\documentclass[journal,twoside]{IEEEtran}
\pdfoutput=1
\usepackage{cite}
\usepackage{amsmath,amssymb,amsfonts}
\usepackage{mathrsfs}
\usepackage{algorithmic}
\usepackage{graphicx}
\usepackage{textcomp}
\usepackage{hyperref}
\newcommand{\argmax}{\operatorname{arg\,max}}

\begin{document}
	
\title{Mining within-trial oscillatory brain dynamics to address the variability of optimized spatial filters}
\author{Andreas Meinel\textsuperscript{1}
	\thanks{$^1$ Brain State Decoding Lab, Cluster of Excellence BrainLinks-BrainTools, Dept. of Computer Science, Albert-Ludwigs-University Freiburg, Germany (e-mail: \{andreas.meinel$|$michael.tangermann\}@blbt.uni-freiburg.de)}, Henrich Kolkhorst\textsuperscript{1,2}\thanks{$^2$ Autonomous Intelligent Systems, Dept. of Computer Science, Albert-Ludwigs-University Freiburg, Germany} and Michael Tangermann\textsuperscript{1}
	\thanks{This work was supported by BrainLinks-BrainTools Cluster of Excellence funded by the German Research Foundation (DFG) by the grants EXC 1086, 387670982 and INST 39/963-1 FUGG as well as by the state of Baden-W\"urttemberg, Germany, through bwHPC. }}

\maketitle

\begin{abstract}
Data-driven spatial filtering algorithms optimize scores such as the contrast between two conditions to extract oscillatory brain signal components. Most machine learning approaches for filter estimation, however, disregard within-trial temporal dynamics and are extremely sensitive to changes in training data and involved hyperparameters. This leads to highly variable solutions and impedes the selection of a suitable candidate for, e.g.,~neurotechnological applications. Fostering component introspection, we propose to embrace this variability by condensing the functional signatures of a large set of oscillatory components into homogeneous clusters, each representing specific within-trial envelope dynamics.

The proposed method is exemplified by and evaluated on a complex hand force task with a rich within-trial structure. Based on electroencephalography data of 18 healthy subjects, we found that the components' distinct temporal envelope dynamics are highly subject-specific. On average, we obtained seven clusters per subject, which were strictly confined regarding their underlying frequency bands. As the analysis method is not limited to a specific spatial filtering algorithm, it could be utilized for a wide range of neurotechnological applications, e.g., to select and monitor functionally relevant features for brain-computer interface protocols in stroke rehabilitation.
\end{abstract}


\section{Introduction}
\label{sec:introduction}
\IEEEPARstart{M}{any} cortical processes on microscopic and macroscopic levels can be described by oscillatory features~\cite{wang_neurophysiological_2010}. When recording rhythmic brain activity by non-invasive imaging techniques such as electroencephalography (EEG), these signals allow to extract task-related information but also enable to describe and characterize the underlying dynamics of cognitive or motor processes~\cite{lopesdasilva_eeg_2013}. While many studies reported that oscillations are correlated with such processes~\cite{dijk:2008}, some studies have even proposed a causal relation~\cite{herrmann_eeg_2016}. 

A state-of-the-art approach to characterize oscillatory activity from EEG recordings is based on verifying the presence of time-locked, frequency-specific envelope modulations. The induced power modulation effect is known as event-related (de)synchronization (ERD/ERS)~\cite{pfurtscheller:1999}. Several approaches to observe and quantify these effects along the spatial, temporal and spectral domain have been proposed~\cite{graimann_quantification_2006,makeig_auditory_1993,makeig_mining_2004}. They allow to describe the cortical regions involved in evoking such power changes~\cite{seeber_eeg_2016,onton_information-based_2006}. 
ERD/ERS effects are mostly studied upon simple tasks, while paradigms with a rich temporal inner structure have been studied less frequently.

In the field of brain-computer interfaces (BCIs), oscillatory features are of specific interest when building a data-driven decoding pipeline using a spatial filtering algorithm~\cite{blankertz:2008}. This model class generally learns a projection for multivariate data to an informative low-dimensional subspace~\cite{de_cheveigne_joint_2014}. Linear methods even allow for a neurophysiological interpretation~\cite{haufe:2014} which supports their usage in closed-loop systems, e.g., in the field of stroke rehabilitation~\cite{soekadar_brainmachine_2015}. 

Unfortunately, data-driven spatial filtering algorithms have to deal with various issues including the low signal-to-noise ratio of high-dimensional EEG recordings, non-stationarities over time and generally small training data sets~\cite{cheveigne_scanning_2015,artoni_relica:_2014}. As a result, even a slight change of the training data can cause a large variability of obtainable oscillatory features~\cite{CasMeiDaeTan15}. In addition, most approaches need to be configured by a set of algorithm-specific hyperparameters such as frequency bands, time intervals or regularization parameters, among others. Every hyperparameter set can result in different features, such that applying a certain set may mean to miss relevant ones. Conversely, modifying the hyperparameters actively bears the chance to detect other, more informative features. 

So far, the data processing pipelines of traditional BCI applications were typically tuned to maximize the \textit{decoding accuracy}. This criterion has been a good choice for learning a model, if the final goal of the BCI system is to gain rapid and precise control. Using performance as an optimization criterion, however, does not consider the functional role of oscillatory features directly~\cite{artoni_relica:_2014}. While a manual assessment of functional relevance may be possible in small hyperparameter spaces, it turns to be impractical if this search space becomes large. Omitting any feature introspection, however, may be a missed opportunity, as details of their ERD/ERS characteristic can provide an equally beneficial criterion from a clinical perspective~\cite{debener_single-trial_2006}. 

Stepping beyond decoding accuracy, current literature reveals only a limited
amount of studies that specifically consider the reliability and physiological plausibility of EEG features (both within and across
sessions/subjects) over a large set of candidate features. 
Generally, clustering
approaches~\cite{artoni_relica:_2014} have been investigated for this purpose, e.g.,~to extract homogeneous groups of spatial 
filters~\cite{krauledat:2008}, to identify EEG
features encoding a similar stimulus
response~\cite{rutkowski_emd_2010}, to partition oscillatory features into
groups of similar spatial, temporal and spectral
properties~\cite{makeig_mining_2004,spadone_k-means_2012,onton_information-based_2006,
	bigdely-shamlo_measure_2013,touryan_common_2016}
or to identify artifactual
components~\cite{rong_magnetoencephalographic_2006}. 
Consequently, it would be desirable to consider both of these aspects, decoding performance and the functional role of features, in a unified approach.

Here, we contribute a data-driven method to identify \textit{informative, reliable and functionally relevant} oscillatory features in the context of a complex visuo-motor hand force task. 
In an offline analysis on data of 18 healthy subjects, we first explore a large configuration space to harvest the variability of oscillatory features derived by a spatial filtering approach. As a functional signature, we exploit the components' rich temporal envelope dynamics  triggered by \textit{multiple} within-trial events, which is then  exploited by a clustering step. Thereby, we condense the large component space to a small set of reliable oscillatory features that reveal homogeneous and distinct envelope dynamics. Hence, we provide a tool for practitioners to assess the functional role of oscillatory features which might enhance the efficiency of closed-loop interaction protocols, e.g., in the context of BCIs for stroke rehabilitation.

\section{Methods and materials}

\subsection{Visuo-motor hand force task}

\subsubsection*{Paradigm description}
\begin{figure}[tb!]
	\centering
	\includegraphics[width=\columnwidth]{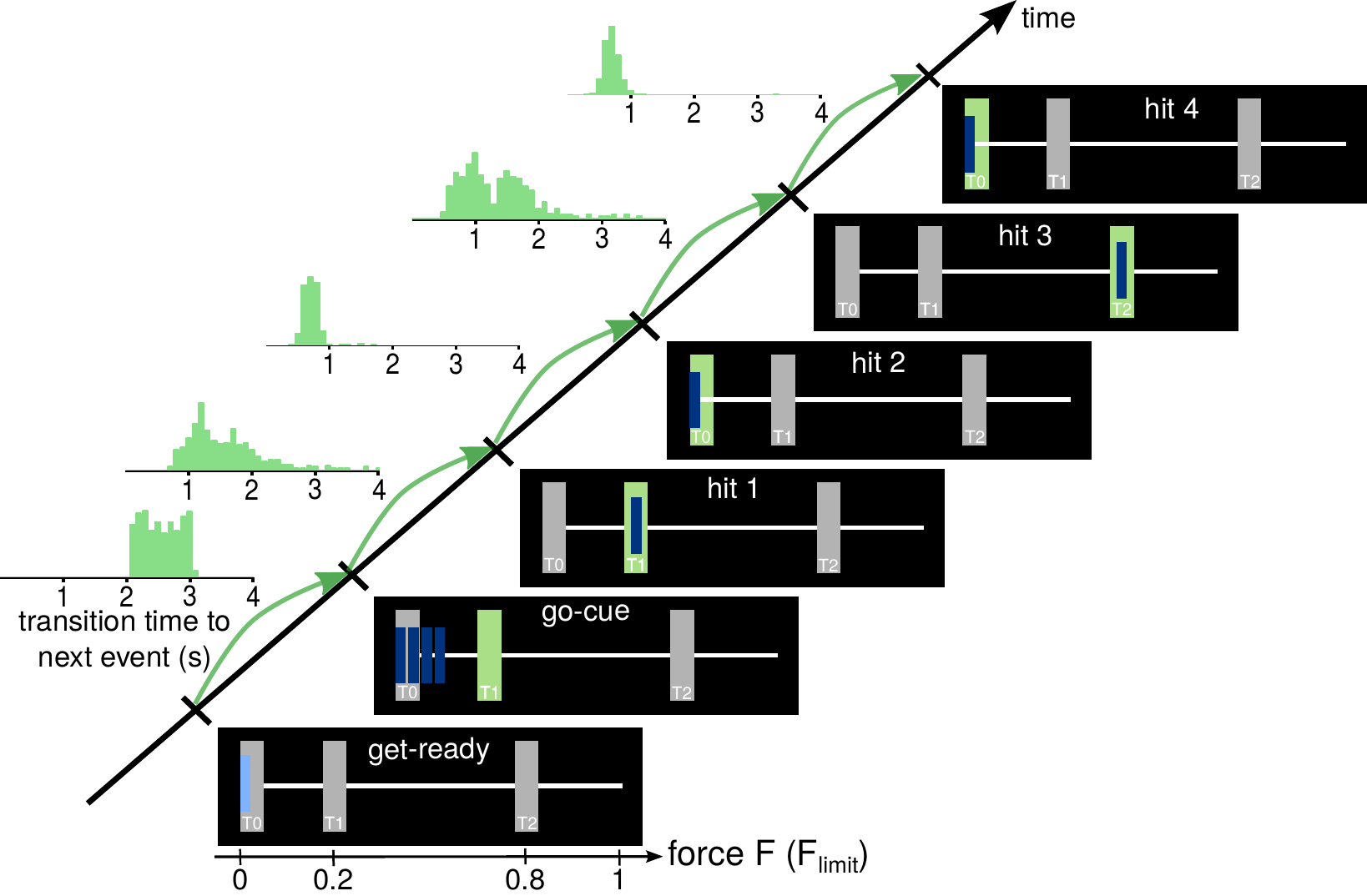}
	\caption{Time course of a single SVIPT trial: For each \textit{within-trial} event, the user feedback on screen is shown (from bottom-left to top-right). Histograms left of the time axis indicate the distributions of transition times between events. The distances between  events on the time axis are not realistically spaced. }
	\label{fig1:svipt_trial_structure}
\end{figure}

Subjects were seated at 80\,cm distance from a 24 inch flat screen and performed a single session of a repetitive visuo-motor task named sequential visual isometric pinch task (SVIPT)~\cite{reis:2009}. They were asked to control a horizontally moving cursor by a force transducer which detects the applied force between thumb and index finger of the non-dominant hand. Throughout a session, brain activity was tracked via EEG for later offline analysis.

A SVIPT trial traverses three stages: 
the appearance of a light blue cursor initiates the \textit{get-ready} phase with duration randomly varied  between 2 and 3\,s. As sketched in Fig.~\ref{fig1:svipt_trial_structure}, an inactive cursor appears on the leftmost border of the target field \textit{T0} (referring to zero force) during this phase, and the force transduction into a horizontal cursor position is inactive. The subject was instructed to fixate the cursor and wait for a \textit{go-cue} which was indicated by switching the cursor's color to dark blue. Starting with the \textit{go-cue} event,  the horizontal cursor position can be controlled by applying force. Its rightmost position at $F_{limit}$ is set to 30\,$\%$ of the user's maximal force.

The \textit{go-cue} marks the running phase of the trial in which the
subject is asked to maneuver the cursor through a sequence of target
fields as fast and as accurately as possible. Thus, overshoots beyond
target fields were to be avoided. A green shading visually indicates
the current target field at any time during the \textit{running} phase
(see Fig.~\ref{fig1:svipt_trial_structure}). Entering a target field
with the cursor, a dwell time of $200\,$ms was required to evoke a
successful hit event. The latter was indicated visually by a switch of
the target field (another green-shaded target field appeared) or by the trial end. Each trial consisted of four hit events (\textit{hit 1} to \textit{hit 4}) and was randomly assigned to one of two target field sequences T1-T0-T2-T0 and T2-T0-T1-T0. 

Overall, data of 18 subjects performing one session of SVIPT with 400 trials are utilized for the presented analysis. Brain activity was acquired by multichannel EEG amplifiers (BrainAmp DC, Brain Products) with a sampling rate of 1\,kHz from 63 passive Ag/AgCl electrodes (EasyCap) placed according to the extended 10-20 system. For details on data preprocessing, see Appx.~\ref{sec:preproc}. The study was approved by the ethics committee in Freiburg. Written informed consent was given by subjects prior to participation. For further details on the study, we refer to~\cite{MeiCasReiTan2016}.

\subsection{Optimized spatial filtering for single-trial EEG analysis}

Let $\mathbf{x}(t) \in \mathbb{R}^{N_{c}}$ describe multivariate EEG data acquired from $N_c$~sensors at time $t$. In addition, let $\mathbf{x}(t)$ be filtered to a narrow frequency band $[f_{0}-\frac{\Delta f}{2},f_{0}+\frac{\Delta f}{2}]$ characterized by a central frequency $f_{0}$ and a bandwidth $\Delta f$. Following the generative model of the EEG~\cite{parra:2005}, a spatial filter $\mathbf{w} \in  \mathbb{R}^{N_{c}}$ represents a linear projection of the sensor space data $\mathbf{x}(t)$ to a one-dimensional source component $\hat{s}(t)=\mathbf{w}^\top \mathbf{x}(t)$. 
Moreover, we assume the data to be segmented into $N_{e}$ single epochs such that $\mathbf{X}(e) \in  \mathbb{R}^{N_{c}\times T_{e}}$ denotes an epoch-wise matrix with $T_{e}$ sample points.

Different optimization criteria for spatial filter estimation can be found. Typical algorithmic solutions for unsupervised scenarios include principal component analysis (PCA), ICA or canonical correlation analysis~\cite{de_cheveigne_joint_2014}. In the context of BCI, supervised algorithms such as common spatial patterns (CSP)~\cite{LotGua2011} or source power comodulation (SPoC)~\cite{DaeMeiHauHoeTanMueNik2013} are state-of-the art approaches, for which also a large number of variants have been proposed.  
Regardless of the specific algorithm, a set of hyperparameters needs to be determined, which influences the spatial filter optimization. Prominent examples are the choice of the training data set, of a subject-specific frequency band or to select components up to a particular rank. In general, choosing different sets from the large search space of possible hyperparameter configurations typically leads to different estimated filters.

In the following, we present a method to identify consistent
spatial filters within a non-trivial hyperparameter space. Even though
this approach can be \textit{utilized for any spatial filtering method}, we exemplarily show and evaluate its application for a regularized version of SPoC named NTik-SPoC~\cite{MeiCasBlaLotTan2018,meinel_a._tikhonov_2017}. This supervised spatial filtering algorithm allows to regress upon a trial-wise target variable given multivariate brain signal data. Details are given in Appx.~\ref{sec:NTik-SPoC}.
As an application scenario, we chose to decode trial-wise reaction
time from multi-channel EEG recordings segmented relative to a time interval close to the SVIPT \textit{go-cue}. A more detailed description can be found in our initial paper~\cite{MeiCasReiTan2016}.

\subsubsection*{Hyperparameter space for oscillatory component analysis}
\label{sec:hyperparamater_space}

For the computation of a single spatial filter $\mathbf{w}$, a set of different hyperparameters is involved (see Figure~\ref{fig2:pipeline_component_mining}). In this paper, we considered the following hyperparameters for the aforementioned SVIPT decoding scenario to explore large variance in the trained decoding models: 

\begin{enumerate}
	\item In the temporal domain, initial data segmentation into $N_{e}$ epochs requires the definition of a time interval $[t_{0},t_{0}+\Delta t]$ with starting time point $t_{0}$ and interval length $\Delta t$. For motor performance decoding in SVIPT, the recorded EEG data were segmented relative to the \textit{go-cue} event with fixed $\Delta t = 1\,s$. Time point $t_{0}$ was chosen among the values $\{-1,-0.75,-0.5\}\,s$.
	\item The frequency domain is characterized by the central frequency $f_{0}$ and bandwidth $\Delta f$. Overall, 45 exponentially increasing and overlapping frequency bands with $f_{0}\in[1,95]$\,Hz and $\Delta f \in [2,10]$\,Hz were evaluated.
	\item The utilized spatial filtering method generally provides a full-rank decomposition. Thus, the rank $k$ can be seen as another hyperparameter for thresholding the decomposition. We considered spatial filters of the first $k=1,...,8$ ranks for further analysis.
	\item NTik-SPoC requires the tuning of the regularization strength $\alpha$. Choosing values in the range $[10^{-8},10^{-3}]$ might allow to outperform non-regularized SPoC wrt.~decoding accuracy~\cite{meinel_a._tikhonov_2017,MeiCasBlaLotTan2018}. In total, 15 discrete, log-scaled values were evaluated of the given range.
	\item Upon each hyperparameter configuration, a 5-fold chronological cross-validation procedure was employed for the calculation of a spatial filter. As many spatial filter optimization problems can be formulated as a generalized eigenvalue problem, a set of filters $\{\mathbf{w}_{k,q}\}$ can be derived from each fold $q$, with $k$ corresponding to the rank in the decomposition. 
\end{enumerate}

In summary, the different hyperparameter configurations span the
configuration space $\Omega=\{(t_{0},f_{0},\Delta f, k, \alpha, q)\}$
with an overall number of $|\Omega|=81,000$ configurations. We will refer to a single configuration by $\omega_{j} \in \Omega$ for $j=1,...,|\Omega|$. It determines a single spatial filter calculation. Hereafter, every spatial filter $\mathbf{w}$ corresponds to a single  configuration $\omega_{j}$.

\subsubsection*{Evaluation scheme for spatial filter computation}

NTik-SPoC was evaluated within a 5-fold chronological cross-validation procedure such that the data is split in train $\{\mathbf{X}_{tr},z_{tr}\}$ and test sets $\{\mathbf{X}_{te},z_{te}\}$. After estimating the labels $z_{est}$ on different test data points $\mathbf{x}_{te}$, the decoding performance of a single filter $\mathbf{w}_{tr}$ can be assessed by the $\operatorname{z-AUC}$ metric~\cite{MeiCasReiTan2016,MeiCasBlaLotTan2018}. This measure characterizes the separability of the estimated labels when comparing them to the known test labels $z_{te}$. In analogy to the area under the ROC curve (AUC), a perfect decoding is reflected by a value of 1, while chance level is allocated to 0.5.
Hereafter, we will omit the train/test subscripts to simplify the notation.

\subsection{Method for mining oscillatory components}
\label{sec:clustering_framework}

\begin{figure}[tb!]
	\centering
	\includegraphics[width=0.4\columnwidth]{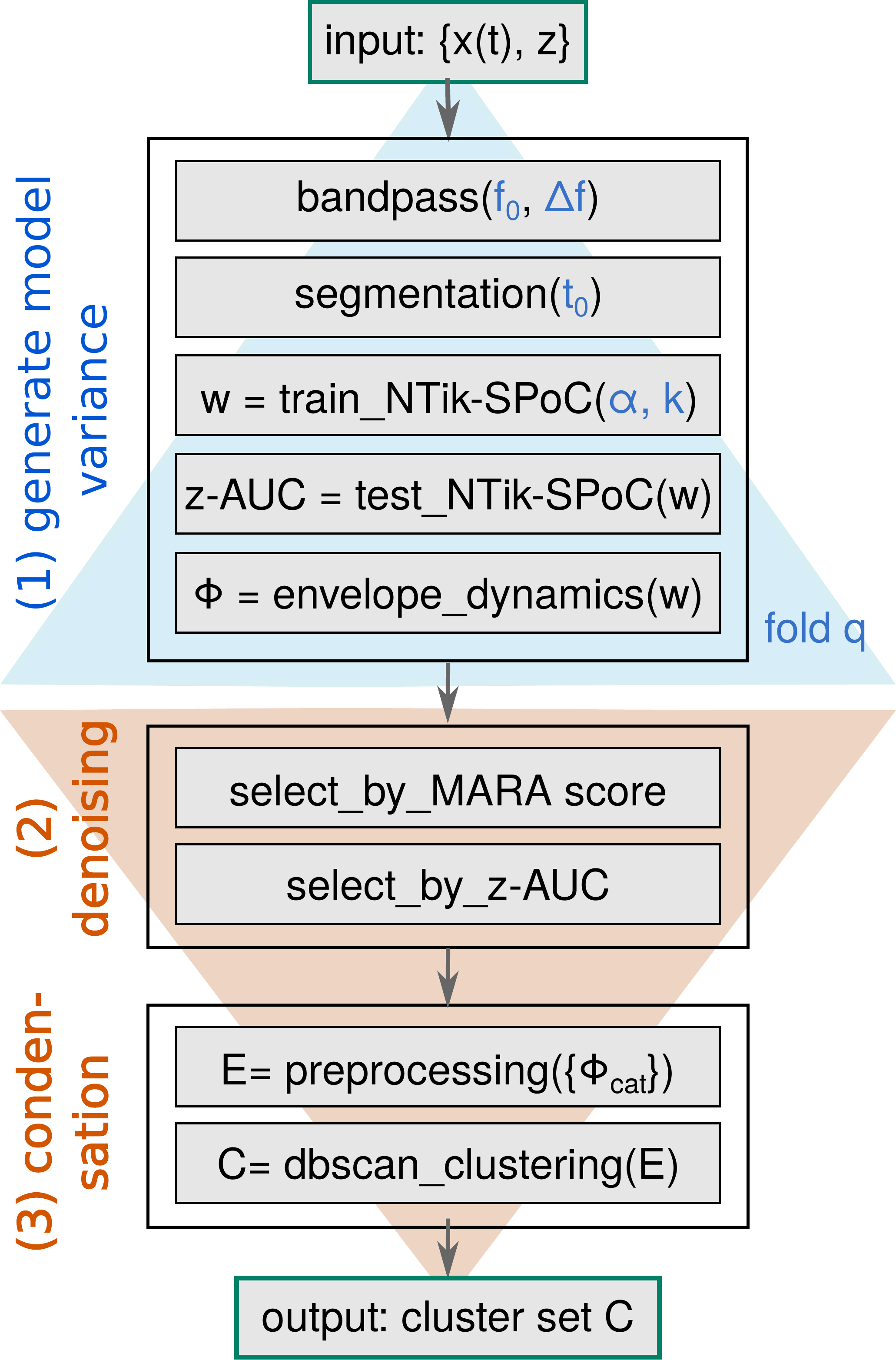}
	\caption{Pseudo-code scheme: subject-specific data $x(t)$ and labels $z$ enter part~(1) to sample oscillatory components in a large hyperparameter space and thus sample variable  spatial filter solutions. In blue, all included hyperparameters for spatial filter $\mathbf{w}$ estimation are highlighted. Part~(2) is a denoising step by reducing the overall configuration space before step~(3) condenses envelope features into clusters.}
	\label{fig2:pipeline_component_mining}
\end{figure}

Our method to identify groups of functionally relevant oscillatory components is depicted in Fig.~\ref{fig2:pipeline_component_mining}. It consists of three steps: 
\noindent
first, oscillatory components from EEG data of a single subject are computed across a large hyperparameter space. This results in a broad variety of spatial filter examples and embraces the variability of the decoding method. Details on the screened configuration space are given in Sec.~\ref{sec:hyperparamater_space}. Second, the large component space is de-noised by restricting the analysis to \textit{reliable} components via selection criteria on single components. Third, for all remaining components the event-related envelope dynamics of each source component is exploited for a clustering step. Therefore, envelope features are preprocessed and condensed. The goal is to finally identify clusters of oscillatory components which reveal distinct and stable envelope dynamics. 

In the following, details on the three steps are given. The code of our method is accessible on GitHub\footnote{\url{https://github.com/bsdlab/func_mining}} and is partially based on the machine learning library \textit{scikit-learn}~\cite{scikit-learn}.

\subsubsection*{Extracting envelope dynamics}
The temporal dynamics of a source component
$\hat{s}$ based on configuration
$\omega_{j}$, given sensor data $\mathbf{x}(t)$, can be derived via
$\hat{s}(t)=\mathbf{w}^\top\,\mathbf{x}(t)$. It requires to bandpass filter the data $\mathbf{x}(t)$ to the same frequency band $[f_{0}-\frac{\Delta f}{2},f_{0}+\frac{\Delta f}{2}]$ on which the spatial filter $\mathbf{w}$ has been trained initially. As an alternative to the variance approximation, the envelope time course $\phi_{j}(t)$ can be estimated by the magnitude of the analytic signal which is given by the Hilbert transformation $\mathscr{H}(\cdot)$ on narrow-band data:
\begin{equation*} \label{eq:hilbert_transform}
\phi_{j}(t)=|\mathscr{H}(\hat{s}(t))|=|\mathscr{H}(\mathbf{w}^\top\,\mathbf{x}(t))|
\end{equation*} 

\begin{figure}[tb!]
	\centering
	\includegraphics[width=\columnwidth]{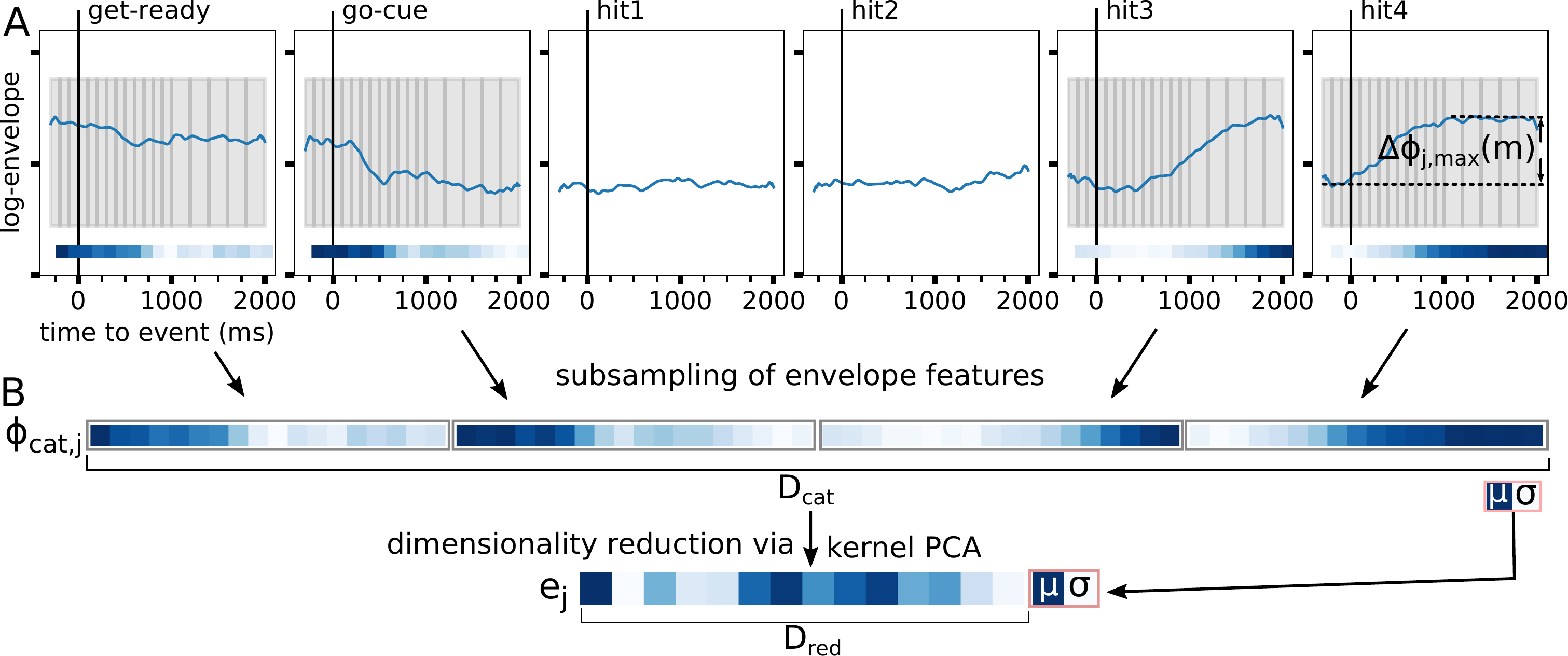}
	\caption{Schematic feature extraction and
		condensation from envelope dynamics of a single
		oscillatory component. (A) The session average log-envelope of
		an oscillatory component was epoched and aligned to the
		various within-trial SVIPT events. In gray, time intervals
		for temporal subsampling of the event-related time series
		are depicted. (B) The subsamples of the four most relevant events 
		were concatenated as well as standardized to zero mean 
		and unit variance as illustrated by the color coding. 
		After dimensionality reduction, ten condensed features plus the two
		standardization features form the final feature vector for clustering.} 
	\label{fig3:envelope_features_clustering}
\end{figure}

For the described SVIPT data, each spatial filter
$\mathbf{w}$ was trained on data segments close to the \textit{go-cue} event. 
However, as depicted in Fig.~\ref{fig1:svipt_trial_structure}, a single SVIPT 
trial also contains a rich inner structure of multiple events $m$ (\textit{get-ready},
\textit{go-cue} and \textit{hit 1} to \textit{hit 4}). We expected that
this within-trial structure is at least partly captured by the temporal
envelope dynamics of single oscillatory components. In other words, 
we tested for the generalization strength of $\mathbf{w}$ on unseen data.
Therefore, the envelope time course of each configuration $j$ was segmented in a
time interval $t\in[-300, 2000]\,$ms relative to each event $m$ and 
averaged across the $N_{e}=400$ epochs. This resulted in
$\phi_j(t,m)$ as exemplarily sketched for one component in Fig.~\ref{fig3:envelope_features_clustering}(A).

\subsubsection*{Denoising clustering input data}

The overall spanned hyperparameter space~$\Omega$ elicits a large
variation of derived spatial filters $\mathbf{w}$
of which a large fraction might be non-stable or of artifactual origin. 
To control for their reliability, the overall component space was 
reduced by deploying two hard selection criteria for the SVIPT decoding scenario: 
\noindent
first, each spatial filter is expected to exceed a robust decoding
performance level above chance. Therefore, each spatial filter is
required to result in a decoding performance above a level of 
$\operatorname{z-AUC}_{min} = 0.6$ on test data. 
The threshold was determined by a group-level
analysis~\cite{MeiCasReiTan2016}. Second, the neural origin of a
spatial filter is verified by applying
MARA~\cite{winkler:2014}, an automatic classification tool to distinguish between
neural and artifactual components, including ocular or muscular activity. 
It returns a posterior artifact probability $p_{art}$ 
for a given single oscillatory component. To restrict the resulting
component space to mostly neural components with a high certainty, we
required a probability of $p_{art}\leq10^{-5}$ for each component.

In summary, the two selection steps reduced the original configuration space $\Omega$ to a subject-specific subset $\Omega_{sel} \subset \Omega$ which comprises all hyperparameter configurations that survived the component selection. Hereafter, the configuration index $j$ refers to the set $\Omega_{sel}$.

\subsubsection*{Features for clustering}

For each configuration $\omega_{j} \in \Omega_{sel}$ the corresponding event-related envelope dynamics $\phi_j(t)$ were extracted as depicted in Fig.~\ref{fig3:envelope_features_clustering}(A). As clustering methods generally depend on the evaluation of a distance metric which is unreliable for high-dimensional spaces, a widely established strategy is then to reduce the dimensionality of the input feature space~\cite{warren_liao_clustering_2005} prior to clustering. Therefore, two steps were taken:

\begin{enumerate}
	\item The logarithmic epoched envelope time series of the four most
	relevant events M=$\{$\textit{get-ready},\,\textit{go-cue},\,\textit{hit 3},\,\textit{hit 4}$\}$ were temporally subsampled to 18 samples per event. This step is schematically shown in Fig.~\ref{fig3:envelope_features_clustering}(A) and resulted in a concatenated feature vector $\phi_{cat,j}=\phi_{cat}(\omega_{j}) \in \mathbb{R}^{D_{cat}}$ with $D_{cat}=72$. After computing mean $\mu$ and standard deviation $\sigma$ of a feature vector $\phi_{cat,j}$, it was standardized to zero mean and unit variance. Overall, applying this procedure to all selected configurations $\omega_{j} \in \Omega_{sel}$, this yielded a subject-specific data set $\Phi=\{\phi_{cat,j}\}$.
	
	\item  A dimensionality reduction step on the concatenated feature set $\Phi$ was performed by kernel PCA to $D_{red}=10$ components. Therefore, a radial basis function kernel with $\gamma=1/D_{red}$ was chosen. The final feature vector $e_{j}$ was composed of the 10 subspace features based on $\Phi$ and two additional features of the standardization step ($\mu$ and $\sigma$) resulting in $e_{j}\in
	\mathbb{R}^{D}$ with $D=12$ dimensions. Overall, a subject-specific data set $E = \{e_{j}\}$ of $|\Omega_{sel}|$ samples was obtained.
	
\end{enumerate}

\subsubsection*{Density-based clustering}

In order to find sets of components with homogeneous envelope dynamics, we aim to group them by searching non-overlapping clusters of components. Assuming that a rich within-trial envelope structure is only expected for a small fraction of all configurations, it may not be necessary to assign each configuration $\omega_{j}$ to a cluster.
Let a clustering of E return a set of disjoint clusters that splits E
into $|C|$ groups with $C=\{c_{k}\}$ and $k=1,...,|C|$. 

For partitioning the data set $E$, the
DBSCAN~\cite{ester_density-based_1996,schubert_dbscan_2017} algorithm
was utilized, which realizes a density-based clustering. DBSCAN groups
dense regions of a data set to clusters by checking for every sample
if: (1) at least $m_{pts}$ other samples are in $\epsilon$ 
range to this sample or (2) at least one neighboring sample in $\epsilon$ 
distance is enclosed. In the first case, the sample is called a \textit{core sample} 
while in the second case it is referred to as a \textit{border point}. 
If none of the two criteria are fulfilled, the sample receives an 
\textit{outlier} label. As such, the density-based definition of a 
cluster requires that each cluster sample reaches \textit{at least one} other 
sample in $\epsilon$ distance.
DBSCAN does not make any assumption on the cluster shape, thus it provides the possibility to
identify non-linearly separable clusters. In this paper, DBSCAN was
evaluated on Euclidean distances $d_{euc}(e_{i},e_{j})$ between samples $e_{i}$ and $e_{j}$ in 
a condensed envelope feature space. 

DBSCAN involves two hyperparameters, $m_{pts}$ and $\epsilon$. Regarding $m_{pts}$, we followed the suggestion of Sander et al.~\cite{sander_density-based_1998} and took the feature dimensionality $D$ into account by setting  $m_{pts}:=2D$.

The choice of $\epsilon$ is the more sensitive parameter, as the number of clusters $|C|$ diminishes strongly for an increased $\epsilon$~\cite{schubert_dbscan_2017}. Overall, we expected that the envelope dynamics would not reveal a rich structure for all hyperparameter configurations $\omega_{j} \in \Omega_{sel}$, e.g., if ERD/ERS effects are not present. Hence, we expected rather large outlier clusters. Thus, a simple maximization of the average silhouette scores S (definition see below) --- which has been used successfully for many other DBSCAN applications --- would most probably be dominated by the outlier class. Instead of using the silhouette score to guide the clustering, we aimed to obtain the maximal number of homogeneous clusters $N_{hom}(\epsilon) = \sum_{c_{k} \in C(\epsilon)} \Theta(\underset{e_{i} \in c_{k}}{\operatorname{min}}(S(e_{i},c_{k}))\geq S_{hom})$ with the unit-step function $\Theta(x) = 1$ for $x \geq 0$ and $\Theta(x) = 0$ for $\,x < 0$. This translates to the following optimization criterion for $\epsilon$:
\begin{equation*}
\epsilon^*=\underset{\epsilon\,\in [\epsilon_{min},\epsilon_{max}]}{\argmax}{N_{hom}(\epsilon)}
\end{equation*}
$N_{hom}$ refers to the total number of clusters, where the sample $e_{i}$ with smallest silhouette $S(e_{i},c_{k})$ of each cluster $c_{k}$ is required to exceed a threshold silhouette score $S_{hom}$. For this paper, a subject-independent threshold of $S_{hom}=0.2$ was set to allow for comparable results such as $|C|$ per subject. This threshold value enforces a slightly smaller within-cluster distance compared to the nearest neighbor distance.
The interval $[\epsilon_{min},\epsilon_{max}]$ was automatically determined by an ordered $k$-distance plot which is based on the k-th nearest neighbor distance (NND) for each sample in E. This procedure was described by~\cite{ester_density-based_1996,schubert_dbscan_2017} together with setting $k=D$. For $\epsilon_{max}$, we detected the first substantial increase of the k-distance plot (starting with smallest distances) by a variance criterion in order to find the end of the ``valley'' of lowest distances. As lower boundary, $\epsilon_{min}$ was determined by the 2nd percentile of the D-th NND distribution. Finally, $\epsilon$ was evaluated at 60 values from the subject-specific interval $[\epsilon_{min},\epsilon_{max}]$.

\subsubsection*{Validation metrics for clustering} \label{sec:validation_metrics}

Given the condensed log-envelope data set of different oscillatory
components, the ground truth cluster labels are unknown. Thus we require a label-free validation metric~\cite{halkidi_clustering_2001,moulavi_density-based_2014} to judge the quality of the clustering outcome. In
this paper, three different categories of validation metrics are
considered~\cite{arbelaitz_extensive_2013}: (1) internal validation
metrics based on features used for the clustering, (2) external
validation metrics which utilize features that have not been used
for the partitioning step and (3) context-specific metrics which are
defined by domain-knowledge given specific characteristics of the
oscillatory component data sets. In the following, for each metric a
reference symbol is given in brackets as well as an arrow that
indicates the direction towards a more preferable clustering (e.g.,
$\uparrow$ denotes ``higher is better''): 

\begin{itemize}
	
	\item Silhouette score $S(c_{k})$ $\uparrow$~\cite{rousseeuw_silhouettes:_1987}: 
	Given an arbitrary sample $e_{i}$ assigned to cluster $c_{k}$, this internal clustering validation score relates the within-cluster similarity $a(e_{i},c_{k})=|c_{k}|^{-1}\sum_{e_{j} \in c_{k}\setminus \{e_{i}\} } d_{euc}(e_{i},e_{j})$ to the nearest neighbor dissimilarity $b(e_{i},c_{k})$:
	\begin{equation*}
	b(e_{i},c_{k})=\underset{c_{l}\in C\backslash c_{k}}{\operatorname{min}}[|c_{l}|^{-1}\sum_{e_{j} \in c_{l}} d_{euc}(e_{i},e_{j})]
	\end{equation*}
	The silhouette score $S(e_{i},c_{k})$ for a sample $e_{i}$ belonging to cluster $c_{k}$ is defined as:
	\begin{equation}
	S(e_{i},c_{k})=\frac{b(e_{i},c_{k})-a(e_{i},c_{k})}{\operatorname{max}[a(e_{i},c_{k}),b(e_{i},c_{k})]} 
	\end{equation}
	It can be  verified easily that $-1\leq S(e_{i},c_{k}) \leq +1$, where a ``perfect'' clustering will return a value of +1. Among other metrics, the silhouette metric has been shown to serve as a reliable \textit{internal} clustering validation method---computed upon features used for the clustering---for various classes of clustering algorithms~\cite{liu_understanding_2010,arbelaitz_extensive_2013}.
	Furthermore, the within-cluster silhouette score $S(c_{k})= |c_{k}|^{-1}\sum_{e_{i} \in c_{k}} S(e_{i},c_{k})$ can be determined with low computational effort.
	
	\item Intra-cluster mean squared error IC-MSE $\downarrow$:
	For the envelope clustering, a number of preprocessing steps were applied in order to reduce the dimensionality of the original time resolved event-related envelopes (see Fig.~\ref{fig3:envelope_features_clustering}). To verify the intra-cluster homogeneity of cluster $c_{k}$ upon the original event-wise log-envelope time series, the event-specific mean $\phi_{avg}(c_{k},t,m)=|c_{k}|^{-1}\sum_{\omega_{j}\in c_{k}}\phi_{j}(t,m)$ for cluster $c_{k}$ was utilized to compute the mean squared error across the full set of events M and time samples~T: 
	\begin{equation}\label{eq:IC-MSE}
	\operatorname{IC-MSE}(c_{k})= \beta^{-1}\sum_{\omega_{j}\in c_{k}}\sum_{m \in M}\sum_{t \in T}(\phi_{j}(t,m)-\phi_{avg})^2
	\end{equation}
	with $\beta = |c_{k}||M||T|$.
	To summarize, IC-MSE corresponds to the intra-cluster envelope variance. It can be seen as an external validation metric as it is based on envelope features which were unseen by the clustering.
	
	\item Intra-cluster pattern heterogeneity ICPH:
	As a context-specific validation metric, the spatial activity pattern $\mathbf{a}_{j}$ for each sample $e_{j}$ of a cluster $c_{k}$ was computed. As a measure of the within-cluster heterogeneity of spatial activity patterns for a cluster $c_{k}$, the cosine angle $\theta$ as defined in ~\cite{MeiCasBlaLotTan2018} between each $\mathbf{a}_{j}$ to a cluster representative pattern $\mathbf{a}^*(c_{k})$ was averaged as follows:
	\begin{equation}
	\operatorname{ICPH}(c_{k}) = |c_{k}|^{-1} \sum_{e_{j} \in c_k}\theta(\mathbf{a}_{j},\mathbf{a}^{*}(c_{k}))	
	\end{equation}
	The representative pattern $\mathbf{a}^*$ was defined by identifying the sample $e^* \in c_{k}$ with minimal Euclidean distance wrt.~clustering features to all other samples of the same cluster.
	
	\item Intra-cluster central frequency variation $\operatorname{std}\big(f_{0}(c_k)\big) \downarrow$: 
	Another context-specific validation was gained via the central frequency hyperparameter $f_{0}$ which is involved in the spatial filter optimization. When capturing the within-cluster variation of $f_{0}$ by $\operatorname{std}\big(f_{0}(c_k)\big)$, this value is expected to be rather small as EEG features are confined with respect to their spectral occurrence.
	
	\item Event-specific maximal envelope difference $\Delta\phi_{max}(m,c_k)$: 
	This metric serves to functionally characterize single clusters by their underlying ERD/ERS dynamics. As exemplarily sketched for \textit{hit 4} in Fig.~\ref{fig3:envelope_features_clustering}, $\Delta \phi_{j,max}(m)$ represents the maximum envelope difference across time within an event~$m$ and for a single configuration $\omega_{j}\in c_{k}$ of a cluster $c_{k}$. Averaging across all cluster configurations reveals the event-specific maximal within-cluster logarithmic envelope differences $\Delta \phi_{max,avg}$:
	\begin{equation}
	\Delta \phi_{max,avg}(m,c_{k})=|c_{k}|^{-1}\sum_{\omega_{j}\in c_{k}}\Delta \phi_{j,max}(m)
	\end{equation} 
	Given a homogeneous cluster, $\Delta \phi_{max,avg}(m) < 0$ refers to an ERD effect for event $m$, while $\Delta \phi_{max,avg}(m) > 0$ describes an ERS effect.
	
\end{itemize}

\subsubsection*{Evaluation scheme for clustering step}

In this paper, we evaluated our proposed methodology on data of 18 subjects. For each subject, the size of the selected configuration space $|\Omega_{sel}|$ was substantially different (see results in Sec.~\ref{sec:statistics}). In order to compare clustering runs across subjects, we randomly sampled $N=2000$ feature vectors $e_{j}$ while keeping their order before entering the clustering. This procedure was repeated twelve times per subject.

\section{Results}

First, we present the most relevant findings for practitioners (e.g.~in the field of BCIs for rehabilitation). Here, representative subject-specific clusters of oscillatory components are described, and we present a way to \textit{functionally} characterize the grouped event-related envelope dynamics. Thereby, we demonstrate the applicability of our method.
Second, we validate and characterize our approach by presenting a
group level analysis across clusterings of all 18 subjects.

\subsection{Findings for the practitioner} 
\label{sec:finding_practitioner}

\subsubsection{Envelope dynamics of clusters} 
\label{sec:exemplary_envelope_dynamics}
\begin{figure*}[htb!]
	\centering
	\includegraphics[width=0.75\textwidth]{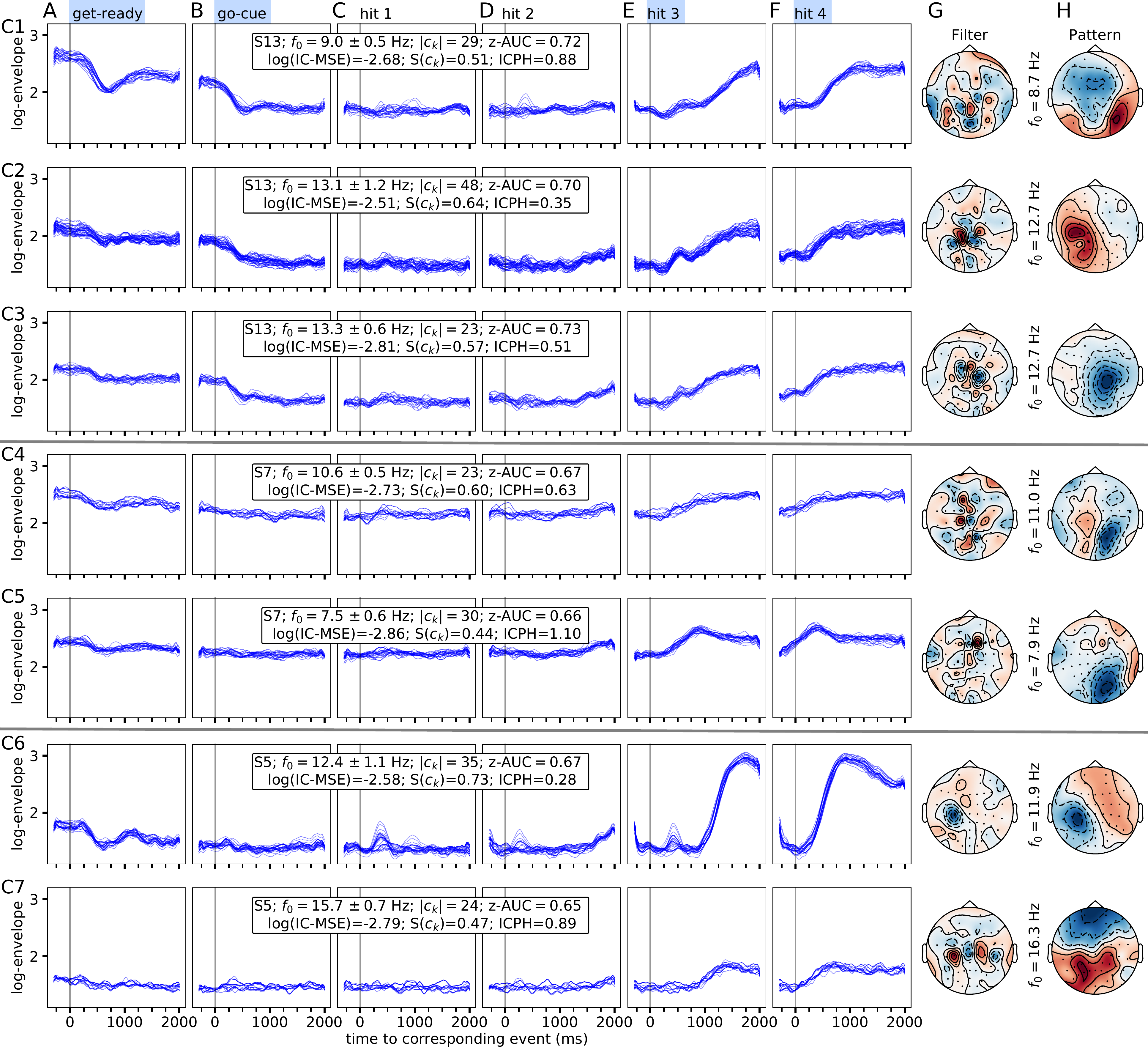}
	\caption{Representative event-related envelope dynamics of all
		hyperparameter configurations for single subject-specific
		clusters $c_k$ reported in rows (C1)--(C7). Columns (A)--(F)
		report the cluster-wise envelope dynamics for within-trial
		SVIPT events, while in (G) the spatial filter and in (H) the
		related activity pattern of cluster representatives (with annotated central frequency) are shown. In all subplots of columns (A)--(F), 
		every blue line refers to the log-envelope dynamics $\phi_j(t,m)$ 
		of one single hyperparameter configuration $\omega_{j} \in
		c_{k}$. Only events shaded in blue were included for the
		clustering step. The text box on top of each row provides
		the subject code, the mean and standard deviation of the
		central frequency across all cluster samples, the
		cluster size, the average decoding performance as well as three validation metrics.  }
	\label{fig4:exemplary_clusters_envelope_dynamics}
\end{figure*}

After applying the described approach, we report the original (non-condensed) within-trial event-related envelope dynamics of subject-specific clusters in Fig.~\ref{fig4:exemplary_clusters_envelope_dynamics}. 
For all examples, the spatial filter and corresponding activity pattern of the cluster representatives --- obtained by selecting the sample with minimal Euclidean distance in the feature space --- are shown in columns (G) and (H).
Rows (C1)--(C7) present different instances of exemplary clusters. They were chosen to represent a broad range of typically observed effects in terms of band-specific amplitude modulations, underlying frequency ranges and cluster homogeneity. Specifically, rows (C1)--(C3) refer to clusters of subject S13, while (C4)--(C5) are gained from subject S7 and (C6)--(C7) correspond to S5. 

Considering the transition times between single events as provided by Fig.~\ref{fig1:svipt_trial_structure}, the time between \textit{hit 1} and \textit{hit 2} as well as between \textit{hit 3} and \textit{hit 4} on average was around 800$\,ms$. Thus, the event-locked envelopes in (C) and (D), respectively (E) and (F), contain overlapping information. 

\noindent Based on Fig.~\ref{fig4:exemplary_clusters_envelope_dynamics}, the following observations can be reported:

first, regarding all shown examplary clusters, the envelope dynamics aligned to the different
within-trial SVIPT events reveal distinct and time-locked ERD or ERS
effects which can be separated well by the clustering approach. For
the cases reported in
Fig.~\ref{fig4:exemplary_clusters_envelope_dynamics}, ERD effects
dominate for \textit{get-ready} and \textit{go-cue} events, while
\textit{hit 3} and \textit{hit 4} elicit ERS effects. 

Second, the displayed examples demonstrate that the event-related envelope dynamics reveal substantially different shapes both within and across subjects. Taking a closer look at the selected cluster instances of S13 with clusters (C1)-(C3) at the \textit{get-ready} event, there are different effects visible. While components grouped into (C2) and (C3) reveal a slight step-like behavior, cluster (C1) comes with a strong ERD followed by an ERS effect. Regarding \textit{hit 4} of (C6) and (C7), the examples for subject S5 nicely illustrate that amplitude differences can be substantially different across clusters/configurations. While clusters (C4) and (C5) of subject S7 are characterized by an ERS effect time-locked to \textit{hit 3}, all remaining examples reveal the ERS at \textit{hit 4}.  

Third, we can report under which conditions our clustering approach works best, e.g., when comparing the cases (C2) and (C3). They nicely demonstrate that smaller cluster
sizes correspond to more homogeneous clusters as visually observable and documented by, e.g., the IC-MSE values.

Fourth, similar silhouette and IC-MSE scores do not directly imply a high pattern homogeneity (low ICPH values), as can be seen comparing clusters (C1) and (C6).

\subsubsection{Functional assessment of clusters}

\begin{figure}[htb!]
	\centering
	\includegraphics[width=\columnwidth]{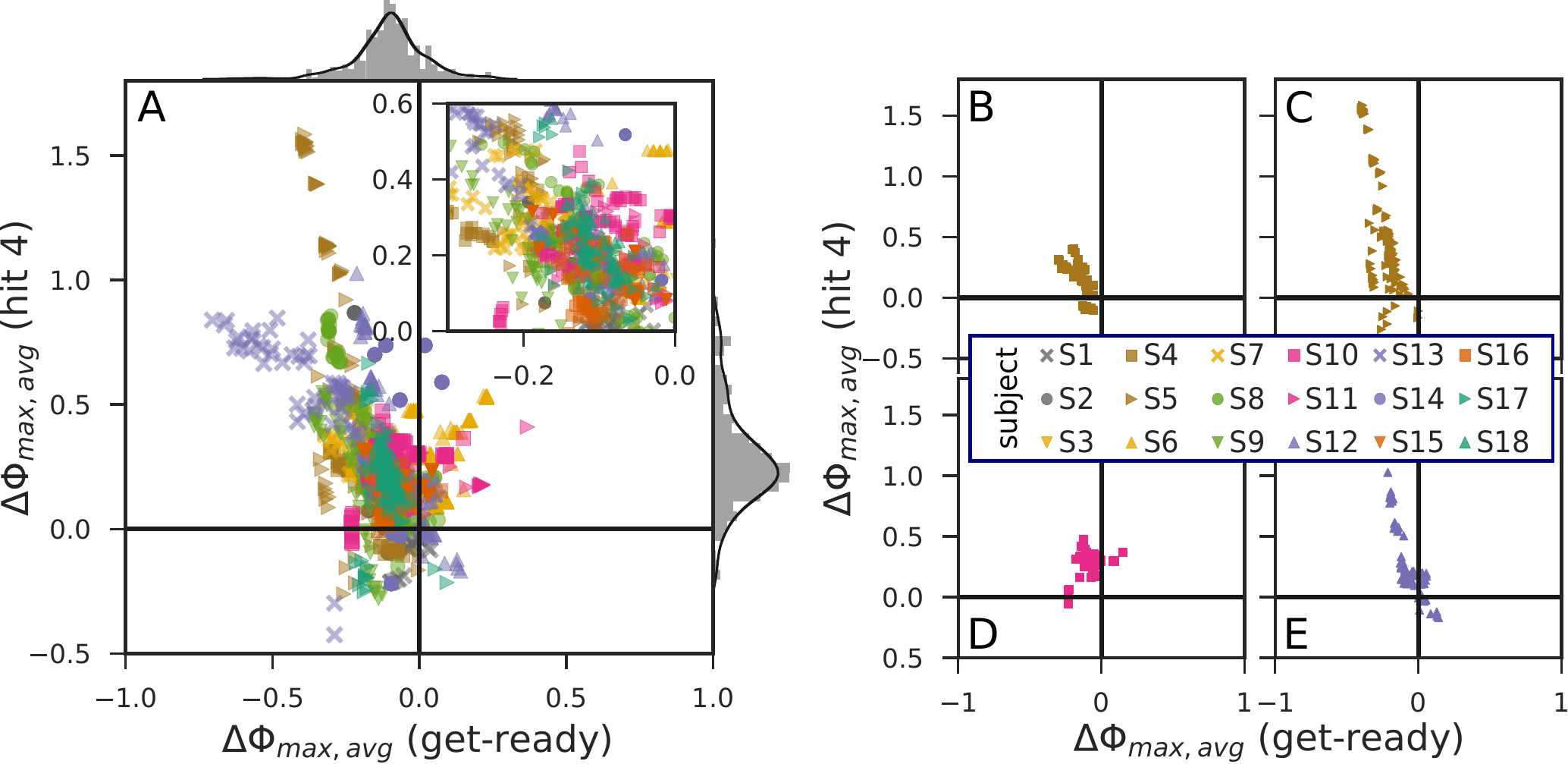}
	\caption{Scatter plots revealing cluster-specific ERD/ERS effects exemplarily for two specific SVIPT events. (A) The plot is based on pooled data from all clusterings across subjects. A single data point refers to an individual cluster and is encoded by its underlying subject. The inset plot shows a magnified version of the central area. Attached to both axes, the corresponding distributions across all displayed configurations are shown. Plots (B)--(E) show exemplary scatter plots for single subjects. }
	\label{fig10:functional_role_clusters}
\end{figure}

Cluster-specific ERD/ERS intensities along the within-trial events indicate 
the functional role of the contained components and thus provide a way to 
characterize single clusters post-hoc. An example is given in
Fig.~\ref{fig10:functional_role_clusters}(A). It exemplarily describes
the logarithmic envelope differences within \textit{get-ready} and
\textit{hit 4} events. 

The distributions on the x- and y-axis reveal ERD effects time-locked to \textit{get-ready} and \textit{go-cue} (not shown here) and a subsequent ERS effect with the last hit event per trial.
Captured by the tails of the distributions reported in Fig.~\ref{fig10:functional_role_clusters}(A), a small fraction of clusters behaves differently and reveals an ERS for ~\textit{get-ready}, which might be caused by remaining artifactual components.

As already stated for the examples given in Fig.~\ref{fig4:exemplary_clusters_envelope_dynamics}, the event- and cluster-specific maximal envelope differences $\Delta \phi_{max,avg}(m,c_{k})$ vary between and even within subjects as exemplarily reported in Fig.~\ref{fig10:functional_role_clusters}(B)--(E). The markers related to a single subject in Fig.~\ref{fig10:functional_role_clusters} are very close or even overlapping for some cases, which is expected due to finding identical clusters over the twelve clustering runs per subject.

\subsubsection*{Group-level cluster statistics}

\label{sec:statistics}
We observed substantial differences across subjects regarding the total
number of available robust oscillatory components that survived the denoising step (see Sec. ~\ref{sec:clustering_framework}). On the grand average across the 18 subjects, $\approx 5,000 \pm 3,900$ configurations corresponding to $6.2\,\pm 4.8\%$ of the initial 81,000 configurations entered the clustering step. As for three subjects less than 2000 configurations were available, only a single DBSCAN run was performed. Across all subjects, we found $7.4\pm3.0$ clusters with at least two clusters for each individual subject.

\subsection{Group-level validation of the method}

For the validation of our proposed approach, we analyzed the final clustering results on a group level in different manners: After inspecting the reliability of obtained clusters, their homogeneity will be investigated by multiple validation metrics. Moreover, a comparison of different validation metrics reveals information on the quality of clustering. The following group level evaluation is based on pooled results across all 18 subjects and the corresponding twelve repetitions.

\subsubsection{Reliability of clustered components}

\begin{figure}[htb!]
	\centering
	\includegraphics[width=0.95\columnwidth]{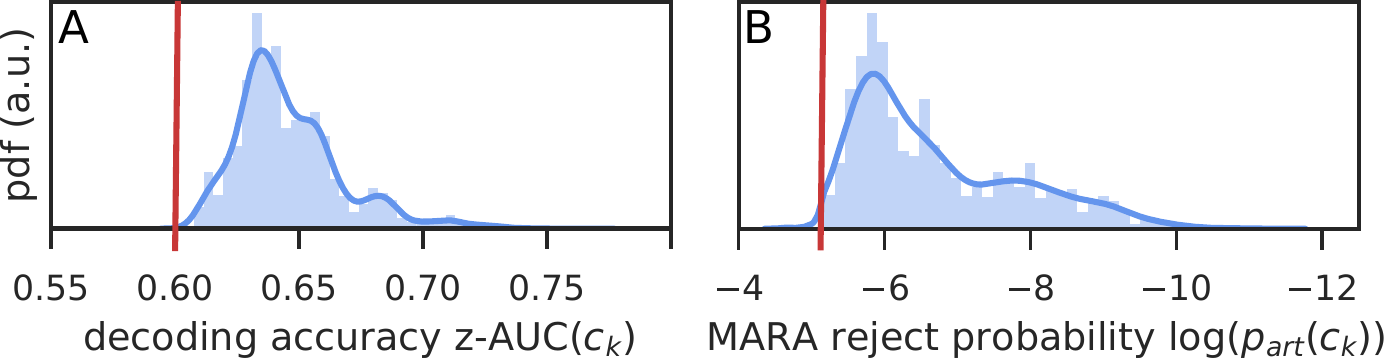}
	\caption{Distribution of the average within-cluster decoding accuracy (A) and the average probability of their artifactual origin (B). The vertical red lines refer to the applied thresholds in the denoising step.}
	\label{fig:clusters_decoding_performance}
\end{figure} 

The denoising step (see Sec.~\ref{sec:clustering_framework}) required all components entering the clustering to provide a minimum decoding accuracy as well as to reveal minimal artifactual contamination. Fig.~\ref{fig:clusters_decoding_performance} nicely demonstrates that the majority of clusters clearly exceed these reliability criteria. The decoding accuracies are in a comparable range as in earlier work on performance decoding for SVIPT~\cite{MeiCasReiTan2016}. 


\subsubsection{Homogeneity of clusters}

\begin{figure}[htb!]
	\centering
	\includegraphics[width=0.95\columnwidth]{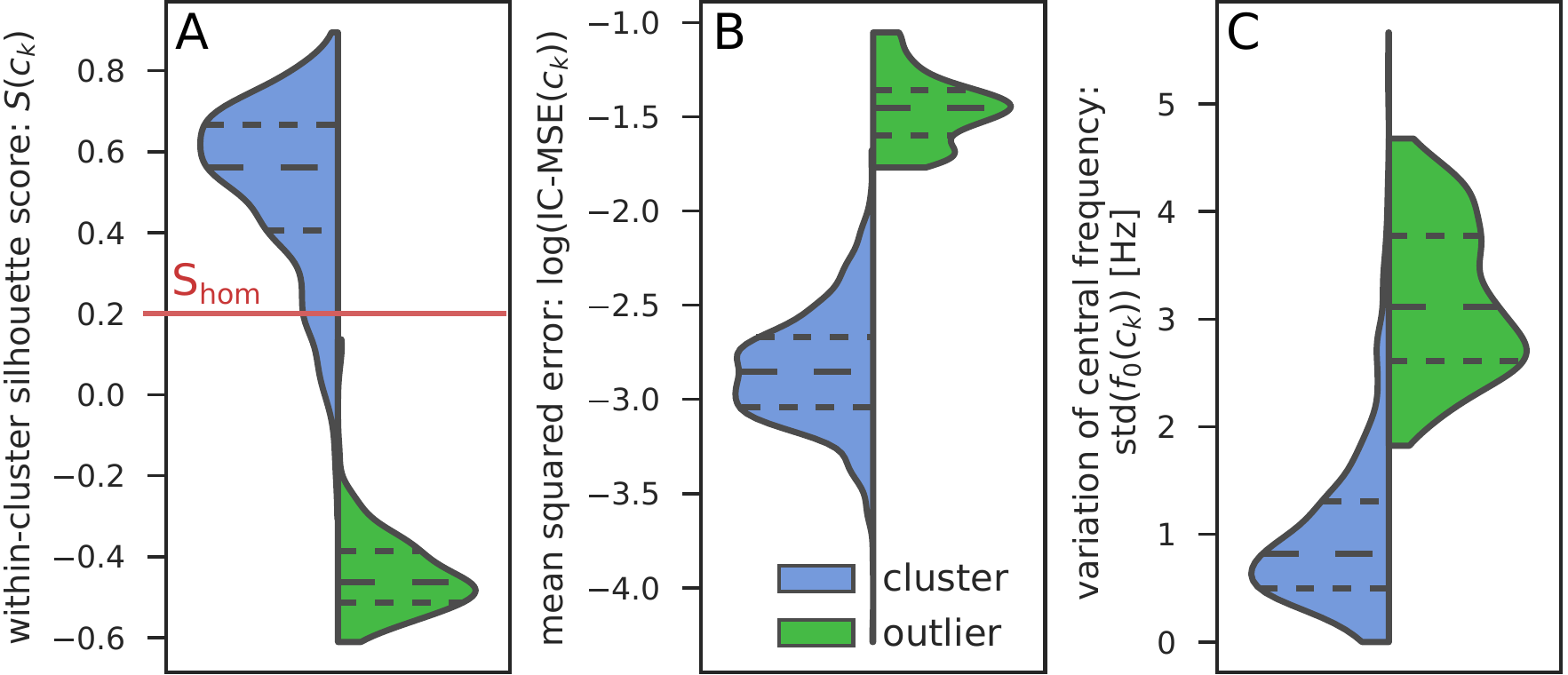}
	\caption{Contrasting distributions of multiple within-cluster evaluation metrics between clusters and outliers, pooled across subjects and twelve repetitions. Internal validation by average silhouette in (A). The horizontal red line corresponds to the global silhouette threshold. As external validation, the log-scaled IC-MSE~(B) is provided. In (C), the within-cluster central frequency variation is shown. For each violon plot, the dashed lines refer to the two quartiles and the median. }
	\label{fig6:cluster_validation_part1}
\end{figure}

DBSCAN separates between cluster and outlier samples. In Fig.~\ref{fig6:cluster_validation_part1}, the distributions for three types of metrics (see Sec.~\ref{sec:validation_metrics}) are contrasted for all identified clusters against detected outlier sets. (A) depicts the distribution for the within-cluster silhouette $S(c_{k})$. 
As enforced by the applied selection criteria for $\epsilon^*$, the 
distribution of $S(c_{k})$ across all clusters is mostly above the 
subject-independent threshold criteria $S_{hom}=0.2$,
while the outlier distribution (blue) is mostly negative. 
In (B), the distribution of the within-cluster IC-MSE is shown. In
accordance with the silhouette distribution in (A) but in reverse
direction, the IC-MSE distribution is strictly shifted towards lower
values compared to the outlier class. 
(C) reports the very confined cluster-specific variance of the central
frequency~$\operatorname{std}\big(f_{0}(c_{k})\big)$ which is consistently below 3\,Hz for clusters. 

\subsubsection{Comparing cluster (validation) metrics}
\begin{figure}
	\centering
	\includegraphics[width=\columnwidth]{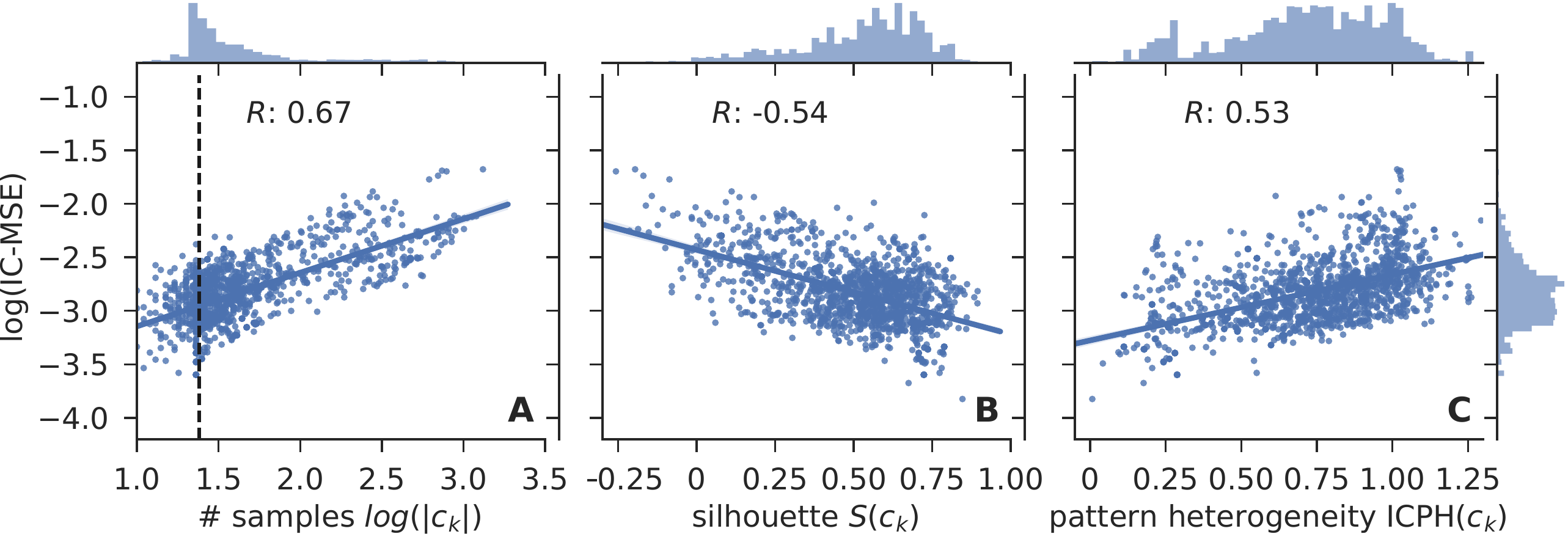}
	\caption{Group-level analysis of cluster (validation)
          metrics. Each data point refers to one single cluster
          resulting from all performed clustering runs across
          subjects. Outliers are not shown here. For all three
          subplots, the log-scaled IC-MSE is shown on the
          y-axis. Three different interactions are shown and the
          corresponding correlation is reported: (A) compares against
          the log-scaled cluster size. The dashed line displays the
          DBSCAN parameter $m_{pts}=24$. In (B), IC-MSE is compared to
          the within-cluster silhouette score and (C) shows the
          interplay with the pattern heterogeneity. For each of the
          four metrics, the corresponding distributions are reported.}
	\label{fig7:cluster_validation_part2}
\end{figure}

The interaction between the external validation metric IC-MSE to the
cluster size $|c_{k}|$ and two other evaluation metrics (see
Sec.~\ref{sec:validation_metrics}) is reported in
Fig.~\ref{fig7:cluster_validation_part2}. In (A), the averaged
envelope homogeneity revealed by the IC-MSE is strongly dependent on
the number of samples $|c_{k}|$ of the respective cluster. The smaller
the cluster size, the more probable it is to find a homogeneous
cluster. The dashed line in (A) reports the DBSCAN parameter
$m_{pts}=24$. If border points are within $\epsilon$-distance of core points of multiple clusters, they will be assigned to only one of them~\cite{ester_density-based_1996}. Thus, we find cluster sizes smaller than $m_{pts}$. Plot (B) shows a negative correlation of the external IC-MSE with the internal silhouette score which was used for DBSCAN optimization. The larger the silhouette value, the smaller the IC-MSE. Similarly in (C), there is a comparable positive correlation between the IC-MSE and the pattern heterogeneity. For small IC-MSE values, a high pattern homogeneity (reflected by low values) is found. However, we also found cases with low IC-MSE but substantial pattern heterogeneity. 

\section{Discussion}

In summary, we propose a method for an informed component selection which addresses and exploits the variability of spatial filter estimates. Our approach identifies groups of oscillatory components that satisfy two optimization criteria in a unified manner: (a) components are required to display a robust decoding performance, and (b) they need to reflect functional relevance for the given experimental task.

Applying these criteria, we demonstrate that the
event-related envelope dynamics of oscillatory components can provide a 
rich source of information, as --- in our example data --- these dynamics are strictly
time-locked to within-trial events of a complex behavioral task. In
addition, we show how this information can be exploited to identify
functionally relevant and reliable oscillatory features from a large
hyperparameter space. Thereby, our data-driven approach is capable to
deal with the noisy character of EEG data. Moreover, it was necessary to design the approach such that it can explicitly cope with rank instabilities as these typically are observed when dealing with eigenvalue decompositions on real-world data sets~\cite{MeiCasReiTan2016}. These instabilities are commonly caused by slight variations of the training data set, or when non-deterministic decoding approaches are utilized.

%

\subsubsection*{Choice of features for clustering}



The clustering step allows to assess the stability and the functional role of oscillatory brain signals.
As discussed for the examples given in
Fig.~\ref{fig4:exemplary_clusters_envelope_dynamics}, exclusively
utilizing the within-trial event-related envelope dynamics for
clustering and no other features, e.g. scalp patterns, turned out to be a suitable
choice for our data scenario. In
Fig.~\ref{fig7:cluster_validation_part2}(C), we reported that single
clusters can contain rather heterogeneous spatial activity
patterns. This could mean that the neural origins of these oscillatory
components differ despite the similarity in ERD/ERS
features~\cite{onton_information-based_2006}. 
Accordingly, Bigdely-Shamlo et al. showed that it might be beneficial to additionally consider 3D dipole locations for clustering, which result from a source reconstruction step~\cite{bigdely-shamlo_measure_2013}. 
We agree on this view, as specifically from a clinical perspective knowledge about functional brain regions could provide added value~\cite{mehrkanoon_upregulation_2016}. However, source reconstruction comes at a price, as results are  sensitive to initial assumptions and raw signal quality, among others~\cite{mahjoory_consistency_2017}. As additional source features would have enlarged our dimensionality, here we decided against including them into our feature vectors. Moreover, the results by Onton et al.~\cite{onton_information-based_2006} support the view that solely event-related dynamics might be sufficient to assess the functional role of oscillatory EEG features. Our observations support this judgment in the context of the investigated behavioral task.

Notably, our results show that the discovered clusters comprise components with highly similar, strictly confined frequency ranges. Overall, most clusters were found to represent oscillations of the alpha- and beta range (not shown). This finding is in accordance with an earlier analysis of the predictive oscillatory EEG features to explain trial-wise motor performance in SVIPT~\cite{MeiCasReiTan2016}.

\subsubsection*{Design choices for the clustering step}

While the literature on clustering of brain signal features is mostly dominated by the k-means approach~\cite{spadone_k-means_2012,touryan_common_2016}, we chose the density-based DBSCAN algorithm for various reasons. First, DBSCAN does not assign each data sample to a cluster. Beside dense clusters, it handles outlier samples without assumptions about the global distribution of outliers. This is beneficial for our data scenario, as we can not expect every configuration to display well-defined envelope dynamics. Second, as DBSCAN does not make an explicit assumption about cluster shapes, it copes well with non-linearly separable clusters. In contrast, k-means is biased towards convex cluster shapes and may not deal well with non-convex shapes. Third, DBSCAN does not require to specify the optimal number of clusters beforehand as a hyperparameter. Given our clustering scenario, this would be difficult to predict.
Motivated by similar arguments, Bigdely-Shamlo et al.~\cite{bigdely-shamlo_measure_2013} utilized an affinity propagation clustering to circumvent the aforementioned shortcomings of k-means. 

Revisiting our full approach, we find several hyperparameters which influence the final clustering. Among others, these comprise the number and width of the time intervals for the temporal subsampling (see Fig.~\ref{fig3:envelope_features_clustering}) as well as the number of components $D_{red}$ obtained from the dimensionality reduction step. The choice of these hyperparameters reveals a trade-off between an adequate temporal resolution of the envelope dynamics and avoiding the curse of dimensionality for large feature dimensions~\cite{aggarwal_surprising_2001}. In principle, the framework could be further optimized by, e.g.,~maximizing the correlation between within-cluster silhouette scores and the IC-MSE scores. 


\subsubsection*{Method applicable for within- and across-subject clusterings}

To allow for a group-level analysis as well as for comparisons between subjects and single subject analyses, we decided to create equally sized data sets before running DBSCAN. For all but three subjects, this requirement lead to a downsampling of components, and we are aware, that we may have omitted informative data especially for subjects with a large component space $\Omega_{sel}$. The repeated downsampling (an ordered random subset selection to $\Omega_{sel}$) typically resulted in quite disjoint subsets, and each of them translated into different clusterings. 

While we restricted the analysis to within-subject clustering, the proposed approach can also be applied to perform across-subject clusterings in the future~\cite{touryan_common_2016}. However, this scenario is more challenging due to strong subject-to-subject variations in brain activity caused by, e.g., anatomical differences which evoke different scalp projections~\cite{bigdely-shamlo_measure_2013,artoni_relica:_2014} of the same functional sources. 

\subsubsection*{Many complex tasks provide a rich inner structure}

One key ingredient of this paper is the exploitation of the rich within-trial event structure of tasks such as SVIPT (see Fig.~\ref{fig1:svipt_trial_structure}). 
Its structure is defined by the sequence of events that occur along a single SVIPT trial. These multiple events offer a large number of choices when it comes to extracting ERD/ERS features: oscillatory activity could have been extracted not only for hit events, but in addition also aligned to error events (such as when the cursor overshoots a target field) or other application-specific events.
The proposed analysis concept should generally be applicable to many complex real-world tasks, as long as they reveal sub steps that can be utilized to define a ``within-task'' event structure. 
Exploiting this additional information, current brain state decoding approaches might gain access to the underlying functional role of features.

\subsubsection*{Identifying functional roles of components on novel data}

Assuming we apply our method on novel data, we propose a simple two-step procedure for identifying functionally relevant components: first, the \textit{reliability} of the identified clusters needs to be verified, e.g.,~using the external IC-MSE metric. A comparison to the distribution reported in Fig.~\ref{fig7:cluster_validation_part2} will enable to judge its reliability. 
Second, one can assess the functional contribution of a cluster's components by investigating the \textit{ERD/ERS characteristics}. Functionally relevant components should reveal amplitude modulations time-locked to at least a few within-trial events (see Fig.~\ref{fig4:exemplary_clusters_envelope_dynamics} and Fig.~\ref{fig10:functional_role_clusters}). 
Along this line, the exact timing of ERD/ERS effects might provide valuable additional information.

\subsubsection*{Expected benefit for targeted closed-loop interaction}

Our method can provide a valuable offline tool to prepare informed
closed-loop interaction protocols. To state an example, we foresee a
potential benefit in the field of BCI protocols for stroke
rehabilitation training~\cite{remsik_review_2016}: first, it can 
allow to introspect the training progress by monitoring underlying 
cortical processes across multiple sessions. As an example, the introspective character of our method may allow contributing to the current debate on judging the role of ipsilesional versus contralesional features to trigger sensory feedback in rehabilitation scenarios after stroke~\cite{dodd_role_2017}.
Second, it may help to increase the efficiency of current BCI systems as it allows for an informed feature selection such that functionally specific BCI feedback can be
realized. 

\section{Conclusion}

We presented a data-driven method for assessing reliable and functionally relevant oscillatory EEG components estimated by a spatial filtering approach. For this purpose, we first embrace the large variability of the generated spatial filters before condensing their functional signatures by a density-based clustering. As a novelty, we make use of within-trial structure for the condensing step. We have evaluated this approach for data of a hand force task, whose inner structure translates into rich temporal dynamics of oscillatory components. 

The proposed approach is not limited to a specific spatial filtering algorithm. By providing introspection about individual, task-related ERD/ERS envelope signatures, we see the method's potential for understanding the functional roles of components. Finally, an informed component selection may increase the efficiency of closed-loop protocols with feedback on oscillatory activity. Overall, these characteristics may be of specific importance for novel protocols in stroke rehabilitation.

\section*{Copyright statement}

\copyright 2019 IEEE.  Personal use of this material is permitted.  Permission from IEEE must be obtained for all other uses, in any current or future media, including reprinting/republishing this material for advertising or promotional purposes, creating new collective works, for resale or redistribution to servers or lists, or reuse of any copyrighted component of this work in other works.

\appendices

\section{Data preprocessing} \label{sec:preproc}
EEG offline preprocessing comprised high-pass filtering the raw
signals at 1\,Hz before low-pass filtering them at 100\,Hz and
sub-sampling them to 500\,Hz sample frequency. For frequency
filtering, linear Butterworth filters of 5th order were
applied. Noisy channels were removed by a two-step
procedure. First, the variance of single epochs and channels was
computed. Based on the pooled statistics, all cases outside the
$[10,90]$ percentiles and also exceeding twice the corresponding
inter-percentile range were registered as outliers. Second,
channels which allocated more than $10\%$ of all outliers and a minimum 
of $5\%$ outliers across epochs were removed. Thereby, $2.9\pm1.9$ 
channels per subject were rejected on average.  
Furthermore, artifact cleaning was done by an independent component analysis (ICA) decomposition on data of the first run of each session. The ICs were rated for artifactual origin with the automated artifact detection framework MARA~\cite{winkler:2014}. Based on MARA's probability ratings, a conservative criterion was applied by removing only up to ten most likely artifactual ICs from the data before projecting the data back into the original sensor space. Only this pre-cleaned data was used in the next steps. For extracting the event-wise envelope information, a supervised spatial filter algorithm was trained. It received data of each channel after segmentation to the interval of $[-1,\,0.5]$\,s relative to the \textit{go-cue} event for each trial. For each of these resulting epochs, a variance and a min-max criterion was applied to remove outliers. For further details see~\cite{MeiCasReiTan2016}.

\section{Regularized source power comodulation (NTik-SPoC)} \label{sec:NTik-SPoC}

NTik-SPoC is a regularized SPoC variant which was introduced and benchmarked in ~\cite{MeiCasBlaLotTan2018}. The supervised spatial filter approach allows to regress upon an epoch-wise univariate behavioral variable $z(e)$. Here, trial-wise SVIPT reaction time---defined by the time interval between \textit{go-cue} presentation and the moment when the cursor left the field T0---was used. 

Given epoched data $\mathbf{X}(e)$, NTik-SPoC optimizes a spatial filter $\mathbf{w}$ such that the epoch-wise source power $p_j(e)=\operatorname{Var}\big(\mathbf{w}^\top
\mathbf{X}(e)\big)$ maximally co-modulates with the variable $z(e)$. This is achieved by solving
$\mathbf{w}^*={\operatorname{argmax}_{\mathbf{w} \in  \mathbb{R}^{N_{c}} }} \,\,
\operatorname{Cov}\big(p_j,z\big)$ constrained by 
$\mathbf{w}^\top \widetilde{\Sigma}_{avg} \mathbf{w}\overset{!}{=}1$
with a shrunk covariance matrix $\widetilde{\Sigma}_{avg}=(1-\alpha)\Sigma_{avg}+\alpha I$. 
This shrinkage operation regularizes the average empirical covariance matrix $\Sigma_{avg}:=\langle\Sigma(e)\rangle$---averaged across all $N_{e}$ epochs---towards the identity matrix~$I\in \mathbb{R}^{N_{c}\times N_{c}}$. The shrinkage intensity is modulated by the regularization parameter $\alpha$. Here,  $\Sigma(e)=(N_{s}-1)^{-1}\mathbf{X}^\top(e) \mathbf{X}(e)$ denotes an epoch-wise covariance matrix. 
The above optimization problem can be formulated as a generalized eigenvalue problem such that solving for $\mathbf{w}$ delivers a full set $\{\mathbf{w}^*_k\}_{k=1,..,N_{c}}$ of $N_{c}$ spatial filters. In this case, the rank $k$ is assigned in descending order of the resulting eigenvalues. NTik-SPoC includes a Tikhonov regularization as described above which in addition is combined with a trace normalization of $\Sigma(e)$ and of the average covariance matrix $\Sigma_{avg}$.

As NTik-SPoC is a linear method, each spatial filter can be visually interpreted by a corresponding average activity pattern $\mathbf{a}=\Sigma_{avg}\mathbf{w}$~\cite{haufe:2014}.

Once a filter $\mathbf{w}_{tr}$ has been computed on training data $\mathbf{X}_{tr}(e)$, it allows to estimate the target variable $z(e)$ on novel, unseen test data $\mathbf{X}_{te}(e)$ by simply estimating the epoch-wise source power $p_j(e)$:
\begin{equation*}  z_{est}(e)=p_j(e)\approx\operatorname{Var}\big(\mathbf{w}_{tr}^\top\mathbf{X}_{te}(e)\big)=\mathbf{w}_{tr}^\top \Sigma_{te}(e) \mathbf{w}_{tr}
\label{eq:spoc_estimate_z}
\end{equation*}

\bibliographystyle{IEEEtran}
\bibliography{general_bib,Component_mining,RegSPoC_paper,SuitAble}

\end{document}